# OCCAM: a flexible, multi-purpose and extendable HPC cluster

M Aldinucci[1,3], S Bagnasco[2,3], S Lusso[2,3], P Pasteris[1,3], S Rabellino[1,3] and S Vallero[2,3]

[1]Dipartimento di Informatica, Università di Torino, corso Svizzera 185, 10149 Torino, Italy

[2]Istituto Nazionale di Fisica Nucleare, via Pietro Giuria 1, 10125 Torino, Italy

[3]Centro di Competenza per il Calcolo Scientifico, Università di Torino, corso Svizzera 185, 10149 Torino, Italy

E-mail: `stefano.bagnasco@to.infn.it`

**Abstract**. The Open Computing Cluster for Advanced data Manipulation (OCCAM) is a multi-purpose flexible HPC cluster designed and operated by a collaboration between the University of Torino and the Sezione di Torino of the Istituto Nazionale di Fisica Nucleare. It is aimed at providing a flexible, reconfigurable and extendable infrastructure to cater to a wide range of different scientific computing use cases, including ones from solid-state chemistry, high-energy physics, computer science, big data analytics, computational biology, genomics and many others. Furthermore, it will serve as a platform for R&D activities on computational technologies themselves, with topics ranging from GPU acceleration to Cloud Computing technologies. A heterogeneous and reconfigurable system like this poses a number of challenges related to the frequency at which heterogeneous hardware resources might change their availability and shareability status, which in turn affect methods and means to allocate, manage, optimize, bill, monitor VMs, containers, virtual farms, jobs, interactive bare-metal sessions, etc. This work describes some of the use cases that prompted the design and construction of the HPC cluster, its architecture and resource provisioning model, along with a first characterization of its performance by some synthetic benchmark tools and a few realistic use-case tests.

## 1. The case for an Open Computing Cluster
Obtaining CPU cycles on an HPC cluster is nowadays relatively simple and sometimes even cheap for academic institutions. However, very large HPC infrastructures may not be suited for some smaller or complex use cases, or for conducting research on HPC technology itself.

The range of use cases proposed by several departments of the University of Torino for its Scientific Computing Competence Centre [1] included ones from solid-state chemistry, high-energy physics, computer science, big data analytics, computational biology, genomics and many others. This heterogeneity calls for different and sometimes conflicting configurations; furthermore, several R&D activities in the field of scientific computing, with topics ranging from GPUs to Cloud Computing technologies, needed a platform to be carried out on. The Centre thus designed and built OCCAM, the Open Computing Cluster for Advanced data Manipulation, described in this paper.

1.1. A wide range of use cases

As mentioned above, OCCAM will not serve only as a traditional HPC cluster, but it will need to be able to accommodate many different applications. In order to design the infrastructure, we focused on three paradigmatic use cases:

- A team from the Chemistry Department of the University of Torino is developing and maintaining CRYSTAL [2], a widely-used software for Ab-initio Solid State Chemistry. The code is developed since the 1970s, and can be applied to the study of any type of crystalline material, with a special focus on the simulation of vibrational spectra. The MPI code does not have huge memory requirements and scales well to thousands of parallel cores, so they need a large number of HPC cores, with little or no need for data access.
- A second team from the Biotechnology and Computer Science Departments is developing CASC [3], a Computational Biology software for Classification Analysis of Single Cell sequencing data. The R-based code is distributed as a set of Docker containers that run in sequence, each using the output of the previous one. Because of the large memory requirements and data access patterns, the software does not scale to more than a few parallel cores, and needs relatively high bandwidth access to data storage.
- A third team at the Economy and Statistics department is analysing atmospheric $NO_2$ concentration using data from air quality monitoring stations and numerical transport models [4]. They use mostly an R-based code to evaluate the uncertainty on forecast curves by applying a bootstrap technique that requires repeated access to a relatively large amount of data, even if the computational power and memory requirements are moderate. Several such use cases do exist, typically R- or python-based code that could run on a single large workstation or a small departmental cluster.

1.2. Bridging the gap

Most computational resources readily available to research teams fall into two broad categories: very large HPC infrastructures, such as the CINECA consortium in Italy, and small departmental clusters. Each of the two models show its limits as soon as the computational needs depart from its ideal use cases. For example, it would often be useful to test code on a smaller, friendlier scale before deploying to very large supercomputers, profiting from a closer interaction with the computing experts who manage the infrastructure. Furthermore, providers of HPC services would usually not allow changes on the hardware or even software configuration, or a lower-level control on the infrastructure and its networks, that may be needed for some studies.

On the opposite end of the scale, local application-specific departmental clusters often suffer from uncertain and sparse funding or lack of professional support, replaced by best-effort work from researchers themselves. Since they are obviously interested in their specific field, and not in scientific computing itself, they usually don't do R&D in HPC and are often very conservative with the tools they use, possibly missing opportunities for a more efficient way of exploiting the computational resources at hand.

An interdepartmental medium-scale Open Computing Cluster, with a multi-disciplinary point of view and a management model more similar to a collaborative research centre than a service-oriented facility, can also be a platform for R&D and innovation, fostering collaboration between Computer Science experts, technology providers, developers and scientific users and thus promoting new technologies in scientific computing practices: new hardware architectures (GPUs, many-core, FPGA,…), software technologies, innovative scheduling tools and strategies, PaaS and SaaS service provisioning models are only some examples of the possible range of activities.

**2. The physical architecture**

The OCCAM supercomputer, in its initial configuration, includes three different types of computing nodes, two storage appliances and two nodes for access and management purposes. All components, both for computing and storage, are connected by three different network links.

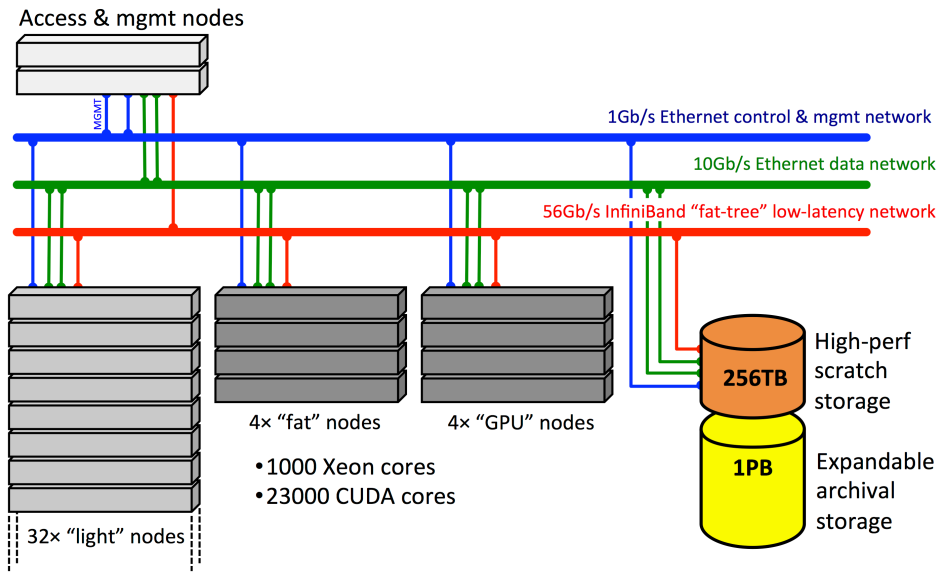

**Figure 1.** Physical architecture of OCCAM

The storage components have different functions: the Archive storage is an expandable general-purpose storage, while the Scratch storage is a high performance appliance for high-bandwidth data access. Even though the Archive storage hardware is fault-tolerant, it is not custodial-grade and there is no backup.

The 1Gb/s Ethernet layer is dedicated to IPMI management of physical components while both the high-performance networks can be used for intra-cluster communication and data access; the fat-tree topology of the IB network guarantees the best performance in parallel throughput between any two components of the system. The configurations of all the components are detailed in tab. 1.

|  | Intel® Xeon® Processors | RAM | GPU | Disk |
|---|---|---|---|---|
| 32 "Light" nodes | 2× E5-2680v3@2.5Ghz | 128GB@2133MHz | Not supported | 400GB SSD |
| 4 "Fat" nodes | 4× E7-4830v3@2.1Ghz | 768GB@1666MHz | Supported on PCI-E Gen3 ×16 | 800GB SSD + 2TB HDD |
| 4 "GPU" nodes | 2× E5-2680v3@2.5Ghz | 128GB@2133MHz | 2× nVidia K40 on PCI-E Gen3 ×16 | 800GB SSD |

|  | Size | Technology | | FS |
|---|---|---|---|---|
| Archive | 768TB | RAID6-equivalent on SATA HDD | | NFS |
| Scratch | 256TB | SDD + HDD hierarchical | | Lustre |

|  | Technology | Topology | Hardware |
|---|---|---|---|
| InfiniBand | 56 Gb/s FDR, non-blocking | Fat-tree | 1x Mellanox SX6036<br>4x Mellanox SX6025 |
| Ethernet | 10Gb/s | Flat | 1x Dell N4000 |
| Ethernet | 1Gb/s | Flat | 1x Dell N3048 |

**Table 1.** Configuration of the system components.

## 3. Benchmarking the resources

Because of the very diverse range of use cases and applications, benchmarking the resources poses a challenge in the selection of the tools to be used, and even on the very metrics to be measured.

### 3.1. Filesystems performance

Because of their different uses, the two storage subsystems were tested for different metrics: the Scratch system was tested for random I/O access and metadata handling performance, while the Archive was tested for sequential I/O only. In the I/O test, the performance of both storage subsystems was measured at the same time, to somehow simulate a realistic working condition.

|          | **Scratch storage** (random I/O, 4k block size) | **Archive storage** (seq. I/O, 64k block size) |
|----------|--------------------------------------------------|-------------------------------------------------|
| Read     | 2026246 kiB/s                                    | 39314 kiB/s                                     |
| Write    | 2025983 kiB/s                                    | 39314 kiB/s                                     |
| Metadata | 10720 sec on 8 nodes                             | N/A                                             |

**Table 2**: storage benchmark results

Flexible IO (*fio*) [5] was used as I/O workload generator, while *mdtest* [6], an MPI-coordinated tool that performs open/stat/close operations on files and directories, was used for metadata performance assessment. The command lines and configuration files used are reported in the Annex, while the values obtained are summarized in tab. 2.

### 3.2. Platform assessment with High-Performance Linpack (HPL)

The LINPACK Benchmarks are a measure of a system's floating point computing power. They measure how fast a computer solves a dense *n* by *n* system of linear equations $Ax = b$, which is a common task in engineering. The latest version of these benchmarks is used to build the TOP500 list, ranking the world's most powerful supercomputers.

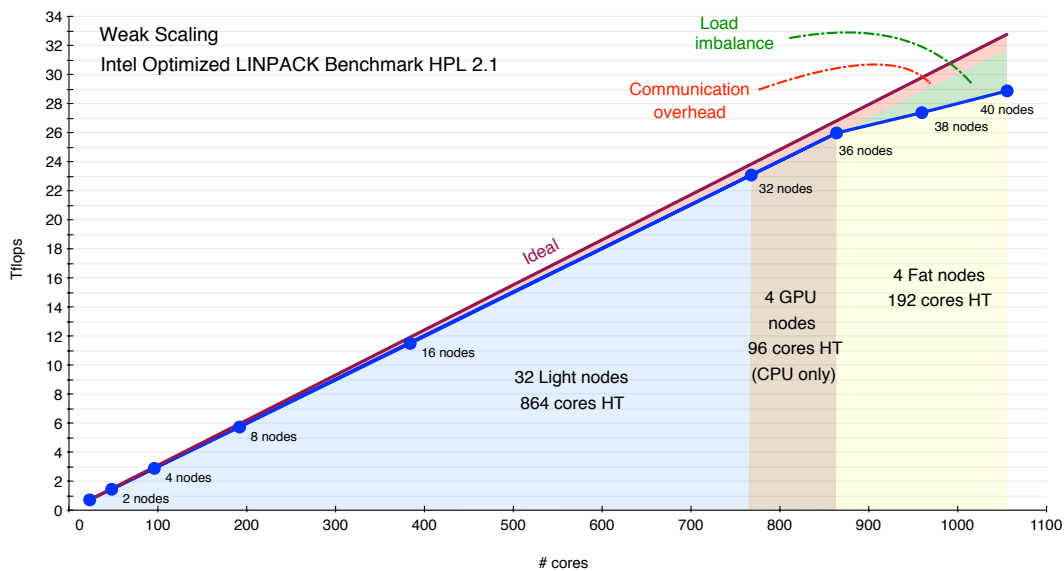

**Figure 2**. Weak scaling results form the High Performance LINPACK benchmark

The Intel Optimized MP LINPACK Benchmark for Clusters [8] is based on modifications and additions to High-Performance LINPACK (HPL) 2.1. It exploits both multiprocessing (via MPI), multi-threading and Advanced Vector Extension (AVX). Specifically, each MPI process queries the node hardware configuration and typically spawn as much threads as real cores. As shown in Fig. 2, the HPL benchmark is able to reach a close to ideal weak scaling on homogenous nodes. Weak scaling measure how efficient an application is when using increasing numbers of parallel processing elements when the problem size assigned to each processing element stays constant (in contrast with strong scaling when the global work stay constant). Notice that the HPL benchmark is an iterative, locked-step application, thus might suffer from load balance issues: at each step all nodes wait for the slowest one. The Intel code might achieve heterogeneous support by distributing the matrix data unequally between the nodes. As all static load balancing methods, its tuning is very time consuming and not always very effective (as shown by the area shaded in red in Fig. 2).

3.3. HEP-SPEC06

Even though it was not a reference use case, High Energy Physics simulation and small dataset analysis may be carried out on OCCAM. The HEP community has developed the HEP-SPEC06 benchmark [9], based on the *all_cpp* subset of the industry standard SPEC® CPU2006 suite, which approximately matches the int/float operations ratio observed in real jobs.

The benchmark was performed on nodes running CentOS 7.3.1611, kernel 3.10.0-514.2.2 and gcc 4.8.5 20150623. In the configuration usually adopted in HEP computing, Hyper-Threading technology was enabled for these tests, so the OS sees twice the number of the physical cores. On Light nodes, the value obtained from 48 concurrent runs is 549.7 HS06; on Fat nodes, 96 runs give 825.6 HS06.

**4. Resource provisioning model and tools**

4.1. The user workflow interface

The main user interface to OCCAM is a web service based on Gitlab [7], a tool that wraps Git with rich user authorization features and several integrated tools, like a Docker image registry, a continuous integration (CI) engine and many other.

Once the user is granted access to the GitLab instance, she can upload the public part of a ssh keypair, that will be used for authentication to hosts, Docker containers or virtual machines allocated by the resource scheduler. Some custom tools then create the user environment: the unix user and group, the home on the archive storage and the openldap entry for keeping information synchronized across nodes.

As containerization through the extensive use of Docker (see below) is a core concept in our execution model, particular attention was given to the Docker private registry integration with GitLab and its CI components to automatically build the images for a project hosted on the system; another interesting feature that will be explored is the gitlab-runner that can enable a user to start computations directly from the gitlab project, eliminating the need of shell login on a node.

4.2. The provisioning and execution infrastructure

The wide range of use cases calls for a great flexibility in the way the Centre provision its users with resources. Instead of simply using a batch system, we choose to borrow some ideas from Cloud Computing technologies, moving the focus from "a user runs some software packages by submitting batch jobs to a queue" to the concept of Computing Applications. A Computing Application is defined by a runtime environment (OS, libraries, software packages, services), resource requirements (large single-image nodes, HPC cluster, GPUs,...) and an execution model (batch jobs, pipelines, interactive access,...). By dynamically partitioning the system, we are able to deploy sandboxed sub-clusters tailored to each application's needs, with the long-term objective of providing a full PaaS-like environment to enable skilled users to build their own applications.

In order to reduce the performance penalty associated with full virtualization, our system is based on the popular Docker software containers [10]. Linux containers, based on chroot, namespaces isolation and kernel resource management features, provide an isolated execution environment for user software; Docker provides a set of tools for easy creation and distribution of computing applications in the form of container images, together with the availability of many off-the-shelf base images.

The use of Linux containers in HPC environments and, more generally, for scientific computing is being explored by several projects, such as the *shifter* project at NERSC [11], or some activities in the INDIGO-DataCloud project [12], from which we take some of the building blocks of our system. Several issues need to be addressed in porting computing applications to containers; for example, popular linear algebra libraries like the Intel® MKL, unless explicitly configured not to do so, will try to scale out threads on a host according to the number of available cores, which may not be the full number reported by the Docker daemon.

In our management model, containers are used for two different purposes:
- by running user-defined images in the system we completely decouple user software management from system management;
- containers are also used to partition the system; each application then comprises a number of containers running on the system, either running services (access, scheduling,...) or executing computing tasks.

However, allowing users to run Docker containers natively on computing nodes poses obvious security concerns. These are currently addressed by either running the containers through *udocker* [13], a tool from INDIGO-DataCloud that, with some limitations, runs Docker container in userspace, or by having the system (instead of the user herself) run the containers in a controlled way.

Even though the final architecture of our system is not fully defined yet, in the current implementation the application-specific subclusters are managed by the Apache Mesos [14] resource manager and controlled by the Marathon scheduler [15], in a configuration again derived by work done in the context of the INDIGO-DataCloud project. The three reference use cases described above are thus mapped onto three different execution environments.

*4.2.1. HPC multi-node workloads.* Such workloads run in on-demand virtual batch farms based on HTCondor [16]. One container acts as an access node running sshd and the HTCondor schedd, another one acts as HTCondor central manager running the collector and negotiator daemons, and a number of executor nodes run the startd daemon and perform the actual computation. The virtual cluster is described through Ansible playbooks and managed by Marathon. If the application needs MPI, the executor containers span the whole node, while if the same infrastructure is used to run single-process batch workloads a container can be assigned only a fraction of a node's resources in order to optimise resource allocation. The clusters are dynamic and can scale out as needed, even though automatic elasticity is not yet available. The full running environment is an adaptation of the HTMesos framework, again from INDIGO-DataCloud [17].

In this case the image to run is derived from a user-defined base image, by adding the relevant clients and configurations. In this development and test phase this is done by hand, but plans are to set up an automated system using the GitLab CI infrastructure.

*4.2.2. Pipeline-like use cases.* Several tools do exist to describe and manage such workflows and can be used together with Mesos to manage the containers. For example, two Common Workflow Language implementations, Toil [18] and Arvados [19], which provide a Mesos backend, are being evaluated. It is interesting to note that while some of the processing steps may easily scale with the number of available processors, some inevitably will not (and some may even be inherently sequential), which poses nontrivial scheduling problems. How to efficiently provision such workflows with containers will be object of future investigation.

*4.2.3. Generic single-node use cases.* Such simpler single-node, single-container applications can be run on compute nodes by way of udocker. In a first implementation, the system deploys an execution container on the assigned node; through some simple commands, the user can instruct the executor to run an image. The executor will then download it to the node and execute it in his userspace through udocker. An interesting possible extension of this use case is using the execution container as a backend for a notebook-style interface like Jupyter [20], as is done in the Swan project at CERN [21].

As a fourth ancillary use case we are also investigating the opportunistic exploitation of unused resources, again using Docker containers and the *plancton* workload management tool [22].

**5. Conclusions and outlook**
Managing a Scientific Computing infrastructure, with widely ranging use cases from HPC to interactive-like workloads, poses a number of problems that we try to address by borrowing some Cloud Computing concepts. The extensive use of Docker containers for software packaging and application isolation allows us to exploit a vast ecosystem of tools, but requires care in the porting of applications, both for security and performance

In Cloud Computing the flexibility of management and possibly increased uptime usually compensate the performance overhead introduced by virtualization; we are confident that this will be the case also in this approach, and an assessment of this assumption will be the next step.

**Acknowledgements**
The OCCAM cluster at the Centro di Competenza sul Calcolo Scientifico of the University of Torino was funded through a contribution by the Compagnia di San Paolo, Italy.

Several tools used in this work were developed by the INDIGO-DataCloud project, funded by the EU Horizon 2020 research and innovation programme under grant agreement RIA 653549.

**Annex**

The command line issued for `mdtest` benchmark is the following:

```
mpirun -f host-mdtest.list -np 32 -ppn 4 /path/to/mdtest -z 5 \
-b 3 -I 2000 -i 64 -w 1048576 -u -d /scratch/md-benchmark
```

where `host-mdtest.list` contains the list of the 8 (light) nodes involved in the test.

Flexible I/O can run in client/server mode. Thus, once started an instance of `fio` server in the involved light nodes, it is possible to launch the benchmark with the following command lines:

```
fio --client=host-archive.list --output=archive.output \
--section=archive --minimal benchmarks.fio

fio --client=host-scratch.list --output=scratch.output \
--section=archive --minimal benchmarks.fio
```

Below the content of the `benchmark.fio` file is listed.

```
[global]
overwrite=1
iodepth=16
ioengine=libaio
direct=1
thread
disable_lat=1
disable_clat=1
disable_slat=1

[scratch]
directory=/scratch/benchmarks
description=Scratch Disk - random read/write test
rw=randrw
bs=4k
rwmixread=50
filename_format=testfile.$filenum.$jobnum.$jobname
size=8G
io_limit=6T
runtime=86400
numjobs=2
group_reporting

[archive]
directory=/archive/benchmarks
description=Archive Disk - sequential read/write test
rw=rw
bs=64k
rwmixread=50
filename_format=testfile.$filenum.$jobnum.$jobname
size=128G
io_limit=32T
runtime=86400
numjobs=2
group_reporting
```